# A Critical Eddington Ratio for X-Shaped Radio Galaxies

David Garofalo
Department of Physics, Kennesaw State University, USA

## Abstract

We derive a quantitative condition for the formation of X-shaped radio galaxies by evaluating the competition between black hole spin evolution and the radiative fading of relic plasma within our previously proposed framework. The simultaneous visibility of two jet axes requires that the timescale for spin evolution across zero, $t_{\rm trans}$, be shorter than the fading timescale of relic radio emission, $t_{\rm fade}$. We estimate the transition timescale as $t_{\rm trans} \approx 5 \times \frac{10^6}{\lambda}$ yr, where $\lambda$ is the Eddington ratio, and derive a visibility timescale $t_{\rm fade} \sim 5$–$20$ Myr based on the evolution of the synchrotron break frequency for typical lobe magnetic fields and redshifts. This leads to a critical Eddington ratio $\lambda_{\rm crit} \sim 0.3$–$1$, above which systems can exhibit X-shaped morphologies. We show that this threshold naturally produces an environmental dependence, as radiatively efficient accretion is more readily sustained in low-density environments, while feedback in rich clusters tends to drive systems toward radiatively inefficient states with a larger fraction of systems having $\lambda \ll \lambda_{\rm crit}$, suppressing XRG formation. We further demonstrate that the observed low fraction of X-shaped radio galaxies ($\sim$ 1–5%) arises from the limited overlap window combined with geometric and detectability effects. These results provide a quantitative and testable extension of our previous model, linking X-shaped morphology to accretion rate and environmental conditions through a simple timescale criterion.

## 1. Introduction

X-shaped radio galaxies (XRGs) are a subclass of radio-loud active galactic nuclei (AGN) characterized by two misaligned pairs of radio structures, a set of high-surface-brightness primary lobes associated with the currently active jet, and a secondary set of diffuse, low-surface-brightness wings oriented along a different axis. These systems provide a unique window into the long-term evolution of relativistic jets and the stability of jet-launching mechanisms over Myr timescales.

Several models have been proposed to explain the origin of the winged morphology. Hydrodynamic backflow models suggest that plasma from the primary lobes is redirected by pressure gradients in the ambient medium, producing secondary structures aligned with the host galaxy's minor axis (e.g., Leahy & Parma 1992; Capetti et al. 2002). Alternatively, rapid reorientation of the jet axis following mergers of supermassive black holes has been invoked, in which a spin-flip leads to a new jet direction while relic emission from the previous episode persists (e.g., Merritt & Ekers 2002; Gopal-Krishna et al. 2012). More gradual jet reorientation scenarios, including disk warping and precession, have also been explored (e.g., Dennett-Thorpe et al. 2002).

Despite these efforts, no single model has successfully explained the combination of observed properties of XRGs, including the symmetry and diffuse nature of the wings and their environmental distribution. XRGs are preferentially found in relatively low-density environments such as field ellipticals and poor groups, and are comparatively rare in rich clusters (e.g., Joshi et al 2019). This environmental dependence presents a challenge for models that rely primarily on strong external confinement or cluster-scale gas dynamics.

In Garofalo, Joshi et al. 2020, it was proposed that XRGs arise naturally within the gap paradigm (Garofalo, Evans & Sambruna 2010) because of black hole spin evolution during accretion. In this framework, mergers can trigger retrograde accretion episodes that produce powerful jets, while continued accretion drives the black hole toward lower spin. The reorientation of the inner accretion disk is expected to occur on the local viscous timescale once the Bardeen–Petterson torque becomes ineffective near zero spin. For radiatively efficient thin disks, the viscous timescale at the relevant inner radii is typically $10^3 - 10^5$yr, depending on the viscosity parameter and disk structure, which is one to three orders of magnitude shorter than the spin-transition timescale of a few Myr derived in this paper. The disk realignment can therefore be regarded as effectively instantaneous on the evolutionary timescale considered here and does not constitute the rate-limiting step in the formation of the new jet. A new jet is then launched along a different axis. Because relic plasma from the previous jet episode persists for a finite time, both jet orientations can be observed simultaneously, producing the characteristic X-shaped morphology.

While this framework provides a natural explanation for the coexistence of multiple jet axes and the preference for low-density environments, it has thus far remained qualitative. The key requirement for the formation of XRGs is that the timescale for spin evolution across zero spin from a value for which the jet is still visible in counterrotation to when a renewed jet emerges in corotation, $t_{\mathrm{trans}}$, be shorter than the timescale over which relic plasma remains visible at radio frequencies, $t_{\mathrm{fade}}$. This competition between intrinsic and radiative timescales has not previously been quantified.

In this work, we derive a quantitative condition for the formation of X-shaped radio galaxies by evaluating this timescale competition. We estimate the transition timescale as a function of accretion rate and derive the fading timescale from synchrotron and inverse Compton losses, focusing on the evolution of the synchrotron break frequency relevant for GHz observations. This leads to the identification of a critical Eddington ratio above which X-shaped morphologies can be produced. We further show that this threshold naturally introduces a strong environmental dependence, as accretion rates are on average regulated differently in low-density and cluster environments.

This paper is organized as follows. In Section 2 we derive the relevant timescales for spin evolution and plasma fading, establish the condition for XRG formation and derive the critical Eddington ratio. In Section 3 we evaluate the expected fraction of XRGs and its dependence on environment. In Section 4 we provide a discussion and conclusion.

2.1 Spin Evolution Timescale Across $a = 0$

In the framework developed in Garofalo, Joshi et al. 2020, the strength of relativistic jets is governed by the efficiency of the Blandford–Znajek mechanism, which extracts rotational energy from the black hole. The power of this mechanism depends sensitively on black hole spin, vanishing as $a \to 0$, where the rotational energy reservoir and frame-dragging effects are minimized (Blandford & Znajek 1977). As a result, there exists a narrow interval around zero spin within which jet production is strongly suppressed. We take this interval to be

$$-0.1 \lesssim a \lesssim 0.1 \qquad (1)$$

corresponding to the transition between retrograde and prograde accretion states in which the jet effectively disappears and subsequently re-emerges (Figure 1). At the Eddington limit ($\lambda = 1$), the black hole crosses this spin interval on a timescale of order

$$t_{\text{trans}}(\lambda = 1) \approx 5 \times 10^6 \text{ yr.} \quad (2)$$

The transition timescale follows from the standard equation governing black-hole growth through thin-disk accretion (e.g. Raine & Thomas 2005). The increase in black-hole mass resulting from the accretion of a rest-mass element dm is given by

$$dM = dm[1 - (2m/3r_{ISCO})^{-1/2}], \quad (3)$$

where $r_{iSCO}$ is the radius of the innermost stable circular orbit. At the same time, the accreting material transfers angular momentum at the ISCO, causing the dimensionless spin parameter *a* to evolve. Integrating the coupled mass-growth and spin-evolution equations across the interval $-0.1 < a < 0.1$ for Eddington-limited accretion yields a characteristic crossing time of approximately $5 \times 10^6$ yrs, which is scale-invariant or mass independent.

We provide a simplified derivation that produces a slight underestimate of the time value as follows. The dimensionless black-hole spin is defined by

$$a = \frac{cL}{GM_{\text{BH}}^2}, \qquad (4)$$

where $L$ is the black-hole angular momentum, $M_{\text{BH}}$is the black-hole mass, $G$ is Newton's gravitational constant, and $c$ is the speed of light. Differentiating Equation (4) gives

$$dL = \frac{GM_{\text{BH}}^2}{c}\, da + \frac{2L}{M_{\text{BH}}}\, dM_{\text{BH}}. \qquad (5)$$

Near $a = 0$, the black-hole angular momentum is itself proportional to $a$, so the second term is higher order in the small quantity $a$ and may be neglected. The spin evolution therefore satisfies

$$dL \simeq \frac{GM_{\mathrm{BH}}^2}{c}\, da. \qquad (6)$$

A mass element $dm$ reaching the inner edge of the accretion disk carries angular momentum

$$dL = dm\, vr, \qquad (7)$$

where $v$ is the orbital velocity at radius $r$. Assuming Newtonian circular motion,

$$v = \left(\frac{GM_{\mathrm{BH}}}{r}\right)^{1/2}. \qquad (8)$$

The inner disk radius varies only weakly across the interval $-0.1 < a < 0.1$, changing from slightly larger than the Schwarzschild ISCO to slightly smaller. We therefore approximate it by the Schwarzschild value,

$$r \simeq 6\frac{GM_{\mathrm{BH}}}{c^2}, \qquad (9)$$

which is accurate to within approximately ten percent over the adopted spin interval. Substituting Equations (8) and (9) into Equation (7) gives

$$dL = dm\left(\frac{GM_{\mathrm{BH}}}{r}\right)^{1/2} r = dm\,\sqrt{6}\,\frac{GM_{\mathrm{BH}}}{c}. \qquad (10)$$

Equating Equations (6) and (10) yields

$$\frac{da}{dt} = \frac{\sqrt{6}}{M_{\mathrm{BH}}}\frac{dm}{dt}. \qquad (11)$$

To determine the accretion rate we assume Eddington-limited accretion. Balancing gravitational attraction against radiation pressure gives

$$\eta c^2 \frac{dm}{dt} = \frac{4\pi G M_{\rm BH} m_p c}{\sigma_T}, \qquad (12)$$

where $\eta$ is the radiative efficiency, $m_p$is the proton mass, and $\sigma_T$ is the Thomson cross section. The corresponding mass accretion rate is

$$\frac{dm}{dt} = \frac{4\pi G m_p M_{\rm BH}}{\eta \sigma_T c}. \qquad (13)$$

Substituting Equation (13) into Equation (11) gives

$$\frac{da}{dt} = \frac{\sqrt{6}\, 4\pi G m_p}{\eta \sigma_T c}, \qquad (14)$$

which is independent of black-hole mass because the factor of $M_{\rm BH}$cancels exactly. This mass independence reflects the fact that both the Eddington accretion rate and the angular momentum required to change the dimensionless spin scale linearly with black-hole mass.

Across the small interval considered here, the radiative efficiency varies only slightly, from approximately 5.7% to 6.7%, so we adopt the Schwarzschild value

$$\eta = 0.06.$$

Equation (11) can therefore be integrated directly,

$$T = \frac{\eta \sigma_T c}{\sqrt{6}\, 4\pi G m_p} \Delta a, \qquad (15)$$

where

$$\Delta a = 0.1 - (-0.1) = 0.2. \qquad (16)$$

Substituting the numerical constants together with $\eta = 0.06$ gives

$$T \simeq 2.2 \times 10^6 \text{ yr}, \qquad (17)$$

for Eddington-limited accretion. More generally, since the spin evolution rate scales linearly with accretion rate, the transition timescale scales inversely with the Eddington ratio $\lambda$, giving

$$t_{\mathrm{trans}} \simeq 2.2 \times 10^{6} \lambda^{-1}\ \mathrm{yr}. \qquad (18)$$

The value of $2.2 \times 10^{6}$yr should be regarded as a lower-limit estimate because the derivation adopts several simplifying assumptions, including a constant ISCO radius and constant radiative efficiency across the interval $-0.1 < a < 0.1$, and neglects higher-order relativistic corrections to the spin evolution. In a full Kerr treatment, both the specific angular momentum and radiative efficiency vary continuously with spin, causing the spin-up rate to decrease slightly as the black hole evolves through the transition. These effects increase the crossing time modestly, so the true transition time is expected to be somewhat longer than the analytic estimate, while remaining of order a few million years. Overall, and since the spin evolution rate scales linearly with the accretion rate, and the transition timescale scales inversely with accretion rate, we parametrize the timescale and adopt a characteristic value of

$$t_{\mathrm{trans}} \simeq 5 \times 10^{6} \lambda^{-1}\ \mathrm{yr}. \qquad (19)$$

The scaling derived here assumes radiatively efficient, geometrically thin-disk accretion and is not intended to describe super-Eddington or slim-disk flows. While slim disks can sustain mass accretion rates exceeding the canonical Eddington limit and would therefore produce even shorter spin-transition times, our model does not prescribe such accretion for the progenitors of X-shaped radio galaxies. Rather, we adopt Eddington-limited thin-disk accretion as the reference case appropriate to the framework considered here. Since super-Eddington accretion would only reduce the transition time further, its inclusion would not weaken the basic timescale criterion but instead make the condition $t_{\mathrm{trans}} < t_{\mathrm{fade}}$easier to satisfy.

The transition time derived above scales linearly with the adopted spin interval,

$$t_{\mathrm{trans}} \propto \Delta a,$$

and therefore, the critical Eddington ratio satisfies

$$\lambda_{\mathrm{crit}} \propto \Delta a.$$

For the fiducial interval adopted in this work, $-0.1 < a < 0.1$ ($\Delta a = 0.2$), the transition time is $5 \times 10^{6}$yr. Halving the interval to $\Delta a = 0.1$ reduces the transition time to $2.5 \times 10^{6}$yr and lowers $\lambda_{\mathrm{crit}}$by a factor of two, whereas increasing the interval to $\Delta a = 0.3$ increases both quantities by 50%. Thus, reasonable variations in the assumed spin interval modify the numerical value of the critical Eddington ratio only by factors of order unity and do not alter the principal conclusion that X-shaped radio galaxies require relatively rapid accretion so that the spin-transition time remains shorter than the fading time of the relic radio plasma.

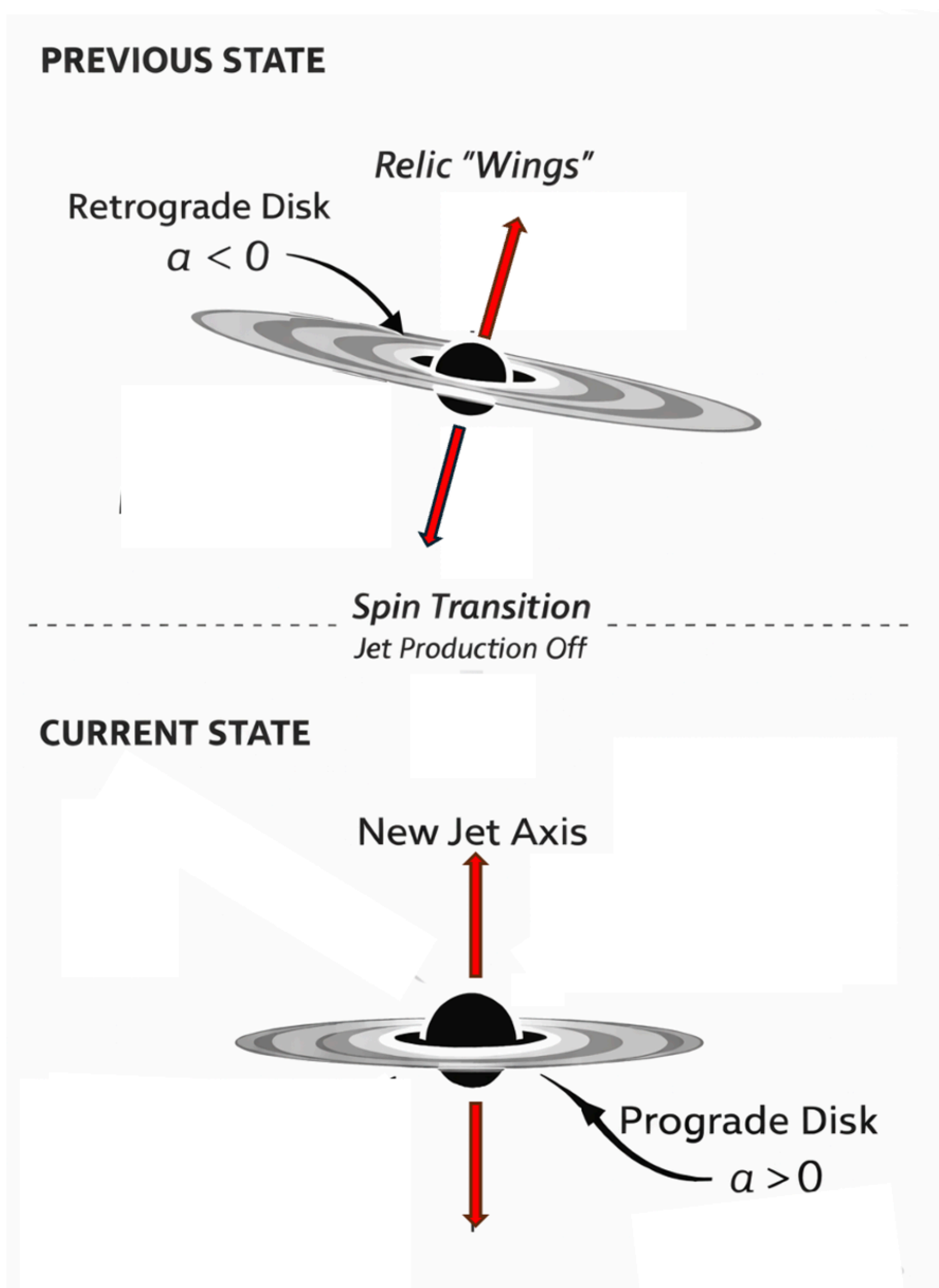

Figure 1: XRG schematic as the result of a counterrotating accretion disk (above) evolving in a few million years into a corotating accretion disk (below).

## 2.2 Fading Timescale of Relic Plasma

The relevant fading timescale is defined as the time over which relic plasma remains detectable at GHz frequencies (e.g., VLA surveys, FIRST, NVSS) following the cessation of jet activity. This is not the total radiative lifetime of relativistic electrons, but rather the timescale over which the synchrotron spectrum evolves such that the break frequency drops below the observing band.

Relativistic electrons lose energy through synchrotron radiation and inverse Compton scattering off the cosmic microwave background (e.g., Kardashev 1962; Pacholczyk 1970; Rybicki & Lightman 1979). The combined energy loss rate behaves according to

$$\frac{d\gamma}{dt} = -\left(\frac{4\sigma_T}{3m_e c}\right)\gamma^2(U_B + U_{rad}),$$

where $U_B = \frac{B^2}{8\pi}$ and $U_{rad}$ includes contributions from ambient photon fields, typically dominated by the cosmic microwave background. Expressing the radiation field as an equivalent magnetic field $B_{\mathrm{CMB}}$, the loss rate scales as

$$d\gamma/dt \propto -\gamma^2(B^2 + B_{\mathrm{CMB}}^2).$$

where $B$ is the magnetic field strength in the lobe and $B_{\mathrm{CMB}}$ represents the equivalent magnetic field of the cosmic microwave background. The CMB energy density $U_{\mathrm{CMB}} = aT^4$can be expressed as an equivalent magnetic field $B_{\mathrm{CMB}} = \sqrt{8\pi U_{\mathrm{CMB}}} = 3.25(1+z)^2$ , enabling a direct comparison between synchrotron and inverse Compton losses. The characteristic synchrotron frequency emitted by electrons of Lorentz factor $\gamma$ is

$$\nu \propto \gamma^2 B.$$

Combining these relations, the synchrotron break frequency evolves as

$$\nu_b(t) \propto \frac{B}{t^2(B^2 + B_{\mathrm{CMB}}^2)^2}$$

The fading timescale is then defined by the condition

$$\nu_b(t_{\mathrm{fade}}) \sim \nu_{\mathrm{obs}}$$

i.e., when the break frequency drops below the observing frequency. For observations at $\nu_{\mathrm{obs}} \sim$ 1GHz and typical parameters for X-shaped radio galaxy wings—magnetic field strengths $B \sim 3$–$5\ \mu$G and redshifts $z \sim 0.1$–$0.3$—the combined loss term is dominated by both synchrotron and inverse Compton contributions, with $B^2 + B_{\mathrm{CMB}}^2 \sim 30$–$60\ (\mu\mathrm{G})^2$. Under these conditions, the resulting fading timescale is

$$t_{\mathrm{fade}} \sim 5 - 20 \text{ Myr}$$

The fading timescale adopted here is intended as an order-of-magnitude estimate appropriate for relic plasma observed at GHz frequencies. The calculation assumes a characteristic lobe magnetic field of a few μG, consistent with values commonly inferred for aged radio lobes and relic structures. We acknowledge that XRG wings may exhibit magnetic fields somewhat below those of active FRII lobes and that processes such as particle re-acceleration or limited injection of fresh relativistic electrons could extend the visibility time. Such effects would increase $t_{\mathrm{fade}}$ and therefore reduce the critical Eddington ratio derived in Section 2.3. Consequently, the values obtained here should be regarded as conservative estimates. The central result of the paper does not depend on the precise choice of magnetic field strength but rather on the existence of a finite relic-plasma visibility time that can be compared with the spin-transition timescale.

A detailed calculation of wing surface-brightness evolution is beyond the scope of this work, as it depends on poorly constrained magnetic fields, particle spectra, and environmental conditions. The essential prediction is that relic wings remain observable for several Myr, permitting overlap with the newly formed jet structure. This represents the timescale over which relic plasma remains visible in GHz radio observations. It is therefore the appropriate quantity to compare with the spin transition timescale when assessing the simultaneous visibility of multiple jet axes

2.3 The Timescale Inequality

The formation of X-shaped radio galaxies requires the simultaneous visibility of two jet axes: relic emission from a previous jet episode and newly formed emission from a reoriented jet. This condition imposes a simple requirement on the relevant timescales

$$t_{\mathrm{trans}} < t_{\mathrm{fade}}$$

where $t_{\mathrm{trans}}$ is the timescale for the black hole to evolve across the spin interval $-0.1 \lesssim a \lesssim 0.1$, during which jet production is suppressed, and $t_{\mathrm{fade}}$ is the timescale over which relic plasma remains visible at GHz frequencies. Substituting the expressions derived in Sections 2.1 and 2.2 gives

$$\frac{5 \times 10^{6}}{\lambda} < (5 - 20) \times 10^{6}$$

This inequality defines the range of accretion rates for which X-shaped morphologies can be produced. Solving for $\lambda$, we obtain

$$\lambda > \frac{5 \times 10^6}{t_{\mathrm{fade}}}$$

For $t_{\mathrm{fade}} \sim$ 5–20 Myr we have

$$\lambda_{\mathrm{crit}} \sim 0.25 - 1.$$

This result shows that X-shaped radio galaxies can only form in systems accreting at a substantial fraction of the Eddington rate. In such systems, the spin transition occurs rapidly enough that relic plasma from the previous jet episode remains visible when the new jet is launched. Conversely, in systems with lower accretion rates ($\lambda \ll \lambda_{\mathrm{crit}}$), the spin evolution timescale becomes longer than the fading timescale, and relic emission disappears before a new jet can form, preventing the appearance of an X-shaped morphology.

3. Fraction of X-Shaped Radio Galaxies and Environmental Dependence

The inequality derived in Section 2.3 implies that X-shaped radio galaxies form only in systems transitioning from counter rotation to corotation while accreting above a critical Eddington ratio, $\lambda_{\mathrm{crit}}$. The observed fraction of XRGs is therefore determined not by a random physical mechanism, but by the fraction of systems satisfying this accretion condition. For systems with $\lambda > \lambda_{\mathrm{crit}}$, the intrinsic fraction of time spent in the X-shaped phase is given by the overlap between relic plasma visibility and the spin transition interval

$$f_{\mathrm{int}}(\lambda) = \frac{t_{\mathrm{fade}} - t_{\mathrm{trans}}}{t_{\mathrm{RL}}} \quad \text{for} \quad \lambda > \lambda_{\mathrm{crit}}$$

and

$$f_{\mathrm{int}}(\lambda) = 0 \ \text{ for } \lambda \leq \lambda_{\mathrm{crit}}$$

where $t_{\mathrm{RL}}$is the total duration of the radio-loud phase. Given $t_{\mathrm{trans}} \sim 5 \times 10^6/\lambda$yr and $t_{\mathrm{fade}} \sim$ 5–20 Myr, the overlap window is modest, typically of order a few Myr. For radio-loud lifetimes $t_{\mathrm{RL}} \sim 10^7$–$10^8$ yr, this yields

$$f_{\mathrm{int}} \sim 0.05 - 0.3$$

The observed fraction of X-shaped radio galaxies is reduced relative to the intrinsic fraction by projection and surface-brightness selection effects. We therefore write

$$f_{\rm obs} \approx f_{\rm int}\, f_{\rm sel},$$

where $f_{\rm sel}$ represents the combined impact of geometric orientation and detectability of low-surface-brightness wings. Based on typical selection biases in radio surveys (e.g., Cheung 2007), $f_{\rm sel}$ is expected to be of order 0.1–0.3. For $f_{\rm int} \sim 0.05$–$0.3$, this yields

$$f_{\rm obs} \sim 1\% - 5\%,$$

consistent with the observed fraction of X-shaped radio galaxies. The key implication of this framework is that the XRG fraction depends on environment through the distribution of Eddington ratios

$$f_{\rm XRG}(E) \propto P(\lambda > \lambda_{\rm crit} \mid E)$$

where $E$ denotes the large-scale environment.

- Low-Density Environments (Field and Poor Groups)

In low-density environments and on average, accretion can proceed in a radiatively efficient mode with moderate to high Eddington ratios. These systems are therefore more likely to satisfy

$$\lambda > \lambda_{\rm crit}$$

and thus to produce X-shaped morphologies. The relatively undisturbed environments also allow relic plasma to remain morphologically coherent over the fading timescale.

- High-Density Environments (Rich Clusters)

In contrast and on average, radio galaxies in rich clusters are subject to strong feedback from powerful jets, which can heat the surrounding medium and reduce the accretion rate. This process promotes an early transition to radiatively inefficient accretion states characterized by

$$\lambda \ll 0.1.$$

In this regime

$$t_{\rm trans} \gg t_{\rm fade}$$

and the relic plasma fades before a new jet can form, preventing the appearance of X-shaped morphologies.

We emphasize that the environmental trends described above are statistical rather than deterministic. Individual radio galaxies in rich clusters may sustain relatively high Eddington ratios for limited periods, just as some field systems may accrete at substantially lower rates. Such objects are not inconsistent with the present framework. Rather, the model predicts that the distribution of Eddington ratios is shifted toward lower values in dense environments owing to more efficient jet feedback and hot-gas accretion, while low-density environments more frequently sustain the radiatively efficient accretion rates required for rapid spin evolution. Consequently, X-shaped radio galaxies are expected to occur preferentially, but not exclusively, in low-density environments, with occasional examples appearing in clusters when sufficiently high accretion rates are maintained.

## 4. Discussion and Conclusions

We have placed our previously proposed interpretation of X-shaped radio galaxies on a quantitative footing by identifying a simple and physically motivated criterion for their formation. The key requirement is that the timescale for black hole spin evolution across the narrow interval $-0.1 \lesssim a \lesssim 0.1$, during which jet production is suppressed, must be shorter than the timescale over which relic plasma remains visible at GHz frequencies. This condition leads directly to a critical Eddington ratio, $\lambda_{\rm crit} \sim 0.25$–$1$, above which X-shaped morphologies can be produced. Reasonable variations in the assumed spin interval change the inferred transition time by factors of order unity, but comparable uncertainties in the accretion rate compensate for these differences, leaving the characteristic timescale in the few-Myr range.

The central implication of this result is that X-shaped radio galaxies are not the outcome of rare or finely tuned events but rather represent a natural phase in the evolution of radio galaxies that satisfy a specific accretion condition. In this framework, the existence of XRGs is governed by a competition between intrinsic evolution of the central engine and the radiative evolution of the surrounding plasma. The resulting inequality introduces a single control parameter—the Eddington ratio—that determines whether a system can exhibit multiple jet axes simultaneously.

This picture naturally explains the observed environmental dependence of XRGs. In low-density environments, where radiatively efficient accretion can be sustained, systems are more likely to satisfy $\lambda > \lambda_{\rm crit}$, allowing rapid spin evolution and the coexistence of relic and newly formed jet structures. In contrast, radio galaxies in rich clusters are more likely to undergo early transitions to radiatively inefficient accretion states due to strong jet feedback, resulting in $\lambda \ll \lambda_{\rm crit}$. In these systems, the spin evolution timescale exceeds the fading timescale, and relic emission disappears before a new jet can form, suppressing the appearance of X-shaped morphologies.

The observed low fraction of XRGs follows naturally from this framework. Even in systems that satisfy the accretion threshold, the overlap window during which both jet axes are visible is limited to a fraction of the total radio-loud lifetime. When combined with geometric projection effects, surface-brightness limitations, and the requirement of a sufficiently large jet reorientation angle, this leads to an observed fraction of order a few percent, consistent with current samples. The present model does not predict the magnitude of the jet reorientation angle. Once the black hole evolves through zero spin, the Bardeen–Petterson alignment torque becomes ineffective, and the orientation of the newly formed accretion disk is determined by the angular momentum of the subsequently accreted gas. The resulting reorientation angle is therefore expected to depend on the merger geometry and post-merger gas dynamics rather than on the spin evolution. Large reorientation angles naturally produce the misaligned jet axes characteristic of X-shaped radio galaxies, whereas small reorientation angles are expected to produce double-double radio galaxies, in which successive jet episodes remain approximately collinear. The distribution of reorientation angles therefore contributes to the observed fraction of X-shaped radio galaxies and is incorporated into the geometric factor introduced in Section 3.

An immediate observational consequence of the present model is that relic wings should be more readily detected at lower radio frequencies. As the synchrotron break frequency decreases owing to radiative losses, the high-frequency emission from the relic plasma fades first, while lower-frequency emission remains detectable for longer periods. Consequently, low-frequency radio surveys should identify a larger population of X-shaped radio galaxies, including systems with faint or undetectable wings at GHz frequencies.

An additional implication of this framework concerns the relative scarcity of high-excitation Fanaroff–Riley type I (FRI HERG) radio galaxies. Observationally, FRI sources are overwhelmingly associated with low-excitation, radiatively inefficient accretion states, while high-excitation systems are typically linked to FRII morphologies. The existence of FRI HERGs is therefore rare and not easily accommodated within standard classifications. In the present model, this apparent anomaly is naturally resolved if FRI HERGs correspond to the newly formed jets that emerge following the spin transition through $a \approx 0$. After reorientation, the jet axis is no longer aligned with the pre-existing evacuated channels produced by the earlier retrograde jet. Instead, the renewed jet propagates into the comparatively dense interstellar medium of the host galaxy (albeit not exclusively), leading to rapid deceleration and disruption of the flow. This interaction produces an FRI-like morphology, even though the system remains in a radiatively efficient, high-excitation accretion state.

In this picture, the FRI morphology is therefore not indicative of intrinsically weak jet production but instead reflects the direct interaction between a newly reoriented jet and the surrounding medium. Because this phase occurs immediately following the spin transition, it coincides with the interval during which relic plasma from the previous jet episode is still visible. As a result, such systems are naturally identified as X-shaped radio galaxies, with the FRI-like structure tracing the new jet axis and the wings corresponding to relic emission from the earlier phase. FRI HERGs are therefore not a distinct or long-lived population, but rather a short-lived transitional phase embedded within the XRG subclass. Their apparent rarity as isolated objects follows directly from the limited duration of the overlap phase. A clear observational prediction is that deep radio imaging of FRI HERG candidates should frequently reveal low-surface-brightness, misaligned relic structures.

More broadly, these results indicate that X-shaped radio galaxies offer a direct observational window into black hole spin evolution on Myr timescales. Their morphology arises naturally from the coupling between accretion-driven changes in spin and the radiative aging of relativistic plasma. In this framework, XRGs mark a transient but inevitable phase in the life cycle of radio galaxies undergoing sustained, high accretion, linking their structure, environments, and apparent subclasses within a unified evolutionary picture. Rather than representing anomalies, X-shaped radio galaxies are the observable imprint of a fundamental competition between spin evolution and radiative losses that governs the behavior of jet-producing black holes.

Acknowledgment

While I only succeeded in part, I thank the anonymous referee for his/her suggestions to make this work more quantitative.